\documentclass[12pt,preprint]{aastex}

\usepackage{natbib}

\usepackage{epsfig,times,lscape,rotating}
\usepackage{color}
\usepackage{amsmath}
\usepackage{amssymb}
\usepackage{txfonts}

\shorttitle{Global disk dynamo}
\shortauthors{Flock et al.}

\begin{document}

\title{Turbulence in weakly-ionized proto-planetary disks}

\author{M. Flock\altaffilmark{1,2}, Th. Henning\altaffilmark{2}, H. Klahr\altaffilmark{2}}

\affil{$^1$CEA Irfu, SAP, Centre de Saclay, 91191 Gif-sur-Yvette, France\\
$^2$Max Planck Institute for Astronomy, K\"onigstuhl 17,
 69117 Heidelberg, Germany}

\begin{abstract}
We investigate the characteristic properties of self-sustained MRI turbulence in low-ionized proto-planetary disks.
We study the transition regime between active and dead-zone, performing 3D global non-ideal MHD simulations of stratified disk covering range of magnetic
Reynolds number between $2700 \lesssim R_m \lesssim 6600$. 
We found converged and saturated MRI turbulence for $R_m \gtrapprox 5000$ with a strength of $\alpha_{SS} \sim 0.01$.
Below $R_m \lesssim 5000$ the MRI starts to decay at the midplane, having Elsasser numbers below one.
We find a transition regime between $3300 \lessapprox R_m \lessapprox 5000$ where the MRI turbulence is still sustained but damped. 
At around $R_m  \lessapprox 3000$ the MRI turbulence decays but could reestablished due to the accumulation of toroidal magnetic field 
or the radial transport of magnetic field from the active region. 
Below $R_m  < 3000$ the MRI cannot be sustained and is decaying. Here hydro-dynamical motions, like density waves dominate.
We observe long-living anti-cyclonic vortices in the transition between dead-zone and active zone.
\end{abstract}

\keywords{accretion discs, magnetohydrodynamics (MHD), dynamo}

\section{Introduction}
The magneto-rotational instability (MRI) is a
candidate for driving turbulence and enabling the accretion of matter onto the central object
\citep{bal91,bal98,haw91}. Dependent on the ionization
degree, the MRI generated turbulence will be reduced, down to a low
turbulence regime, called the dead zone
\citep{bla94,jin96,san00}. Various studies showed that a
certain level of resistivity suppresses MRI activity \citep{fle00,san02I,san02II,fle03,inu05,tur07,tur10,tur08}. 
One of the most important dimensionless number, characterizing the coupling between gas and magnetic fields, 
is the magnetic Reynolds number $R_m$ which
relates the timescale of magnetic advection to magnetic diffusion. We consider here the Ohmic diffusion term which is most important at the midplane.
Up to now, there is no clear prescription for which values of $R_m$ the
MRI turbulence is sustained in disks. \citet{fle00} found sustained MRI with a zero-net flux magnetic field for 
$R_m > 10^{4}$. 
A recent study by \cite{sim11}, using stratified local box simulations,
investigate the turbulence level for different values of the magnetic
Reynolds and Prandtl number. They found a so-called low state, a state where
turbulence is partly suppressed, but sustained due to a dynamo process. 
In addition, they predict a critical Reynolds numbers $R_m^{crit}$ in the range between $3200
< R_m^{crit} < 6000$.  A similar region was investigated by \citet{ois11} in which they found $R_m^{crit} \sim 3000$. This critical Reynolds number is important to model the surface density of active layers in proto-planetary disk as it was recently done by 
\citet{mar12}. \\  
In our study, we will search for the critical magnetic Reynolds number in global zero-net flux stratified MRI simulations. Here the MRI turbulence criterion, Elsasser number $v_{Az}^2 /  (\eta  \Omega) $, should become unity. In contrast to the magnetic Reynolds number, the Elsasser number gives one clear threshold independent of the magnetic geometry or the stratification. We will also investigate the hydrodynamical motions which become important in the dead-zone region \citep{ois09}.
We concentrate on the magnetic Reynolds
number regime proposed by \citet{sim11} and \citet{ois11}. 
For our simulations we use only explicit resistivity.
\citet{ois11} found out that well ionized MRI turbulence scales independently of  $Pm$ if $Rm > Rm^{crit}$.
In addition, as the molecular viscosity is very small in proto-planetary disks we expect Prandtl numbers of $Pm << 1$ and we focus on this low Prandtl number regime \footnote{Here the hydrodynamic viscosity is small compared to
the magnetic diffusivity $\nu << \eta $}.
In this paper we will first describe the numerical and physical setup. Then we will present the results, discussion and the conclusion.
\section{Setup}
The initial conditions for density, pressure and azimuthal velocity 
follow hydrostatic equilibrium. We set the density $\rho$ to
\begin{equation}\rho = \rho_{0}  R^{-3/2}\exp{}\Bigg(\frac{\sin{(\theta)}-1}{(H/R)^2}\Bigg) \end{equation}
with $\rho_{0} = 1.0$, the scale height to radius $\rm H/R = c_0 = 0.07$, $R = r \sin{(\theta)}$.
The pressure follows locally an isothermal equation of state: 
$P = c_{s}^2\rho$ with the sound speed $\rm c_{s} = c_0/\sqrt{R}$.
The azimuthal velocity is set to \begin{equation}V_{\phi} = \sqrt{\frac{1}{r}}\Bigg(1- \frac{2.5}{\sin(\theta)}c^2_0 \Bigg).\end{equation}
The initial velocities $V_{r}$ and $V_{\theta}$ are set to a white noise
perturbation amplitude of $V_{r,\theta}^{Init} = 10^{-4} c_{s}$.
We start the simulation with a pure toroidal magnetic seed field with constant plasma beta
$\beta = 2P / B^{2} = 25$.
To obtain a range of magnetic Reynolds number we keep the magnetic dissipation value constant in the disk. 
We use three different values of $\eta$, $\eta_1=2\cdot 10^{-6} AU^2/yr$, $\eta_2=2.6\cdot 10^{-6} AU^2/yr$ 
and $\eta_3=3.2\cdot 10^{-6} AU^2/yr$. 
\begin{equation}
R_m = \frac{c_s H}{\eta} = 2450 \cdot \left ( \frac{H/R}{0.07} \right )^2 \cdot \left ( \frac{R}{1AU} \right )^{0.5} / \left ( \frac{\eta}{2\cdot 10^{-6}} \right ) 
\end{equation}
To estimate the numerical magnetic Reynolds number having MRI turbulence simulations 
we run also ideal MHD simulations with different resolutions as reference.
The radial domain extends from 1 to 10 AU.
The $\theta$ domain covers $\pm$ 4.3 disk scale heights, or $\theta = \pi/2 \pm 0.3$.
For the azimuthal domain we use $2\pi$ for the L models and $\pi/2$ for the H models.
We use a uniform grid in spherical coordinates. 
 Models L have a resolution of $N_r= 384$, $N_\theta=192$ , $N_\phi=768$ and $N_\phi=384$ for the H models.\\
 All models resolve the radial scale height with 9 to 22 grid cells for the inner and outer radius. The vertical scale height is resolved by 22 grid points. In models L the azimuthal scale height is resolved by 9 grid cells. The H models have a higher resolution of 17 per scale height in the azimuth. Their are calculated with the FARGO MHD to reduce even more the numerical dissipation \citep{mig12}. The simulation models are summarized in table 1. We note that model $L^1$ is special. Here the numerical dissipation cannot be neglected. By comparing with the results of H models, 
the $L^1$ model shows magnetic Reynolds number below the value used in $H^3$. This model establishes a large dead-zone region. Here hydro-dynamical motions become important.
\begin{table}
\begin{tabular}{llllll}
Model name & $N_r  N_\theta  N_\phi$ & $\Delta r : \Delta \theta : \Delta phi$ & FARGO & Rm \\
\hline
\hline
$\rm L^{Ideal}$ & $384$x$192$x$768$ & 1-10 : 0.6 : $2\pi$ & NO & Ideal \\
$\rm L^{Ideal FARGO}$ & $384$x$192$x$768$ & 1-10 : 0.6 : $2\pi$ & YES & Ideal \\
$\rm H^{Ideal FARGO}$ & $384$x$192$x$384$ & 1-10 : 0.6 : $\pi/2$ & YES & Ideal \\
$\rm H^{1}$ & $384$x$192$x$384$ & 1-10 : 0.6 : $\pi/2$ & YES & 4300-6300 \\
$\rm H^{2}$ & $384$x$192$x$384$ & 1-10 : 0.6 : $\pi/2$ & YES & 3300-4800 \\
$\rm H^{3}$ & $384$x$192$x$384$ & 1-10 : 0.6 : $\pi/2$ & YES & 2700-4300 \\
$\rm L^{1}$ & $384$x$192$x$768$ & 1-10 : 0.6 : $2\pi$ & NO & $<$3000* \\
\hline
\end{tabular}
\label{mri-t}
\caption{From left to right: model name, resolution, domain size, FARGO-MHD, range of magnetic Reynolds number. In model $L^{1}$ the used explicit resistivity is not resolved and we estimated the magnetic Reynolds number.}
\end{table}
Buffer zones extent from 1 to 2 AU as well as from 9 to 10 AU.
In the buffer zones we use a linearly increasing resistivity (up to $\eta = 10^{-3}$) reaching the boundary. This damps 
the magnetic field fluctuations and suppresses boundary interactions. 
For our analysis we use the range between 3 and 8 AU, which is not affected by the buffer zones. 
Our outflow boundary condition projects the radial gradients
in density, pressure and azimuthal velocity into the radial boundary and the
vertical gradients in density and pressure at the $\theta$ boundary. 
For all runs we employ the second order scheme in
the PLUTO code with the HLLD Riemann solver \citep{miy05}, piece-wise linear
reconstruction and $2^{nd}$ order Runge Kutta time integration. 
We treat the induction equation with the "Constrained Transport" (CT) 
method in combination with the upwind CT method described in
\citet{gar05}, using explicit resistivity. 
A more detailed description of the physical setup can be found in \citet{flo11}.

\begin{figure}
\vspace{-1cm}
\psfig{figure=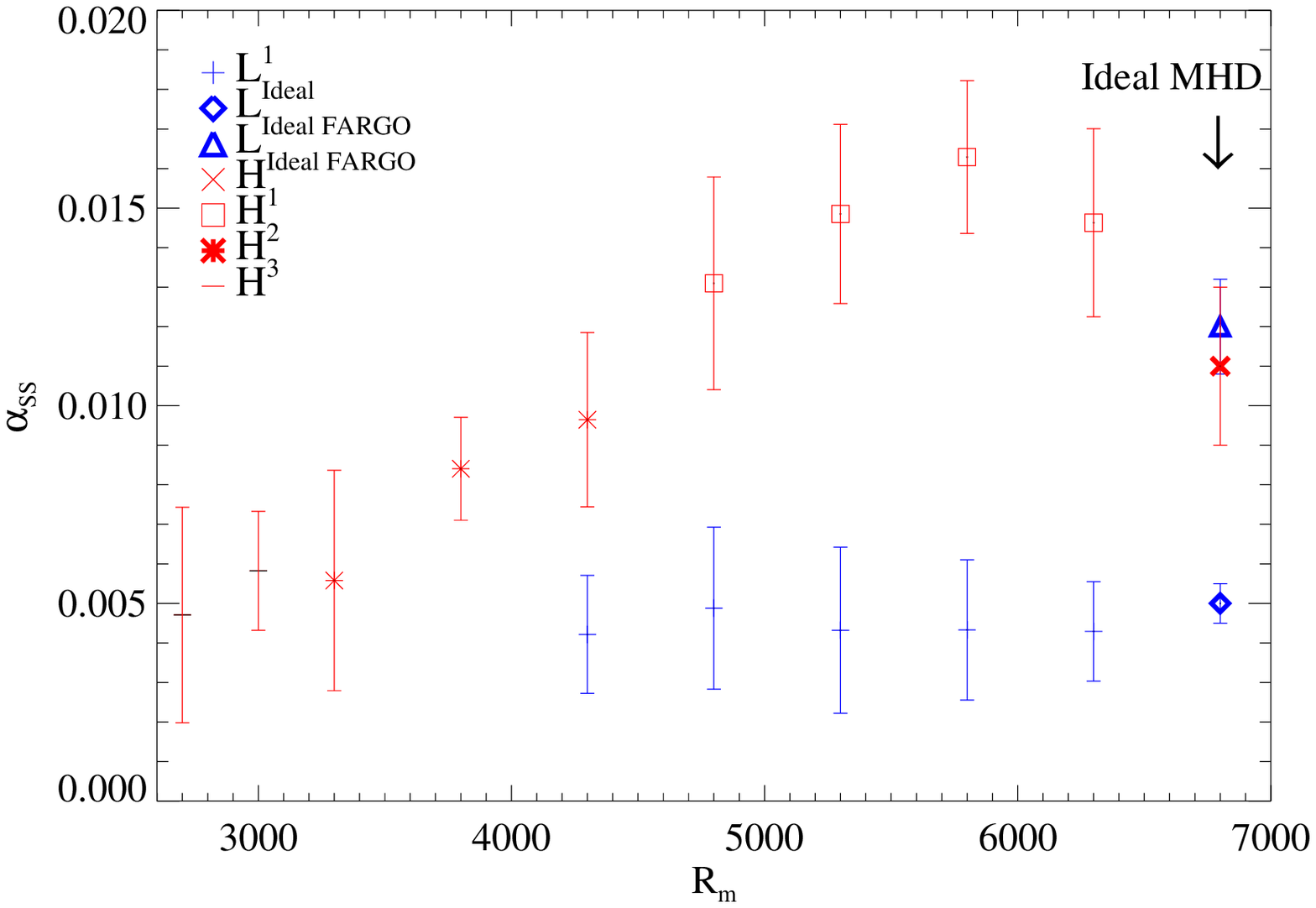,scale=0.57}\\
\vspace{-0.5cm}
\psfig{figure=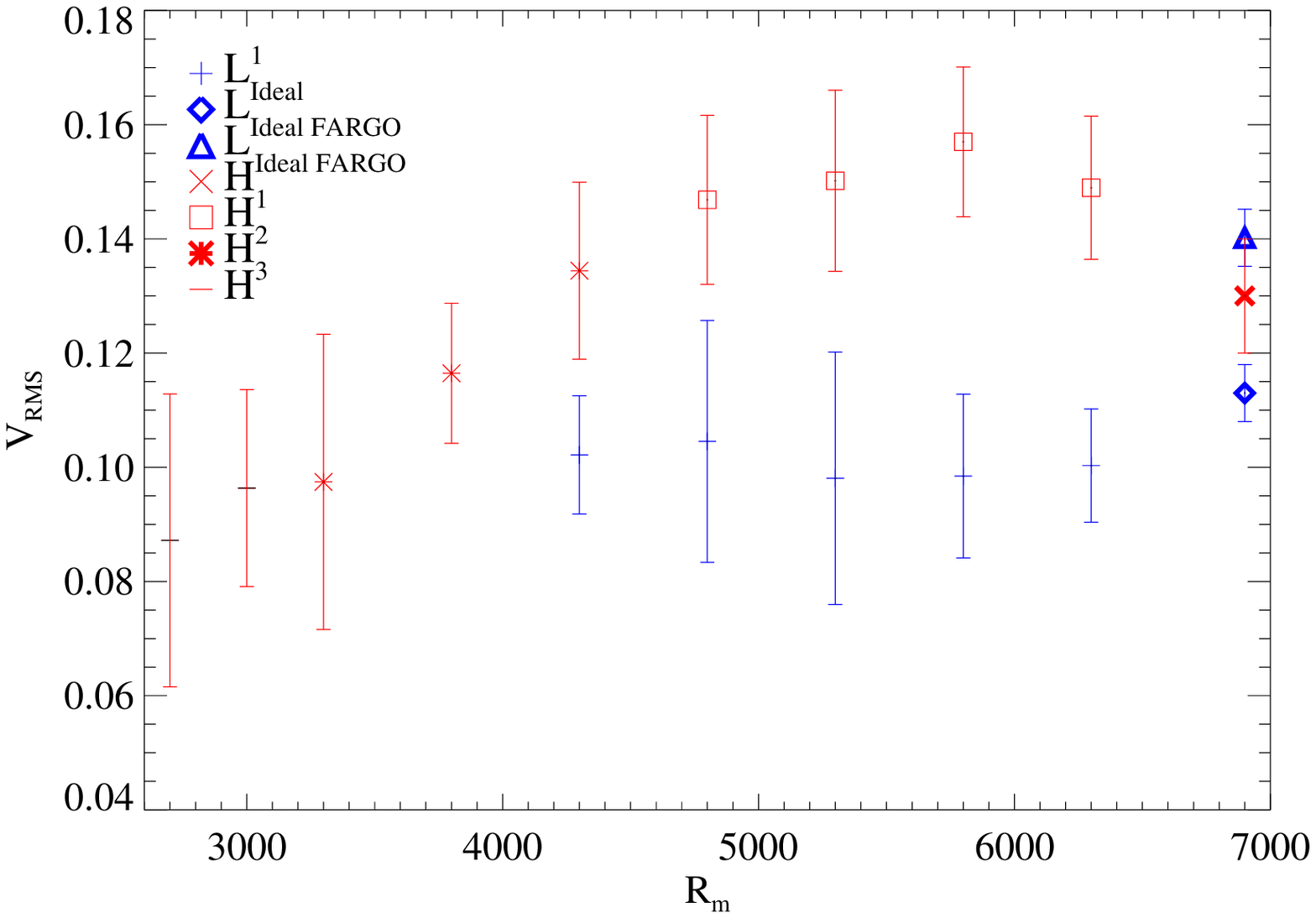,scale=0.57}\\
\vspace{-0.5cm}
\psfig{figure=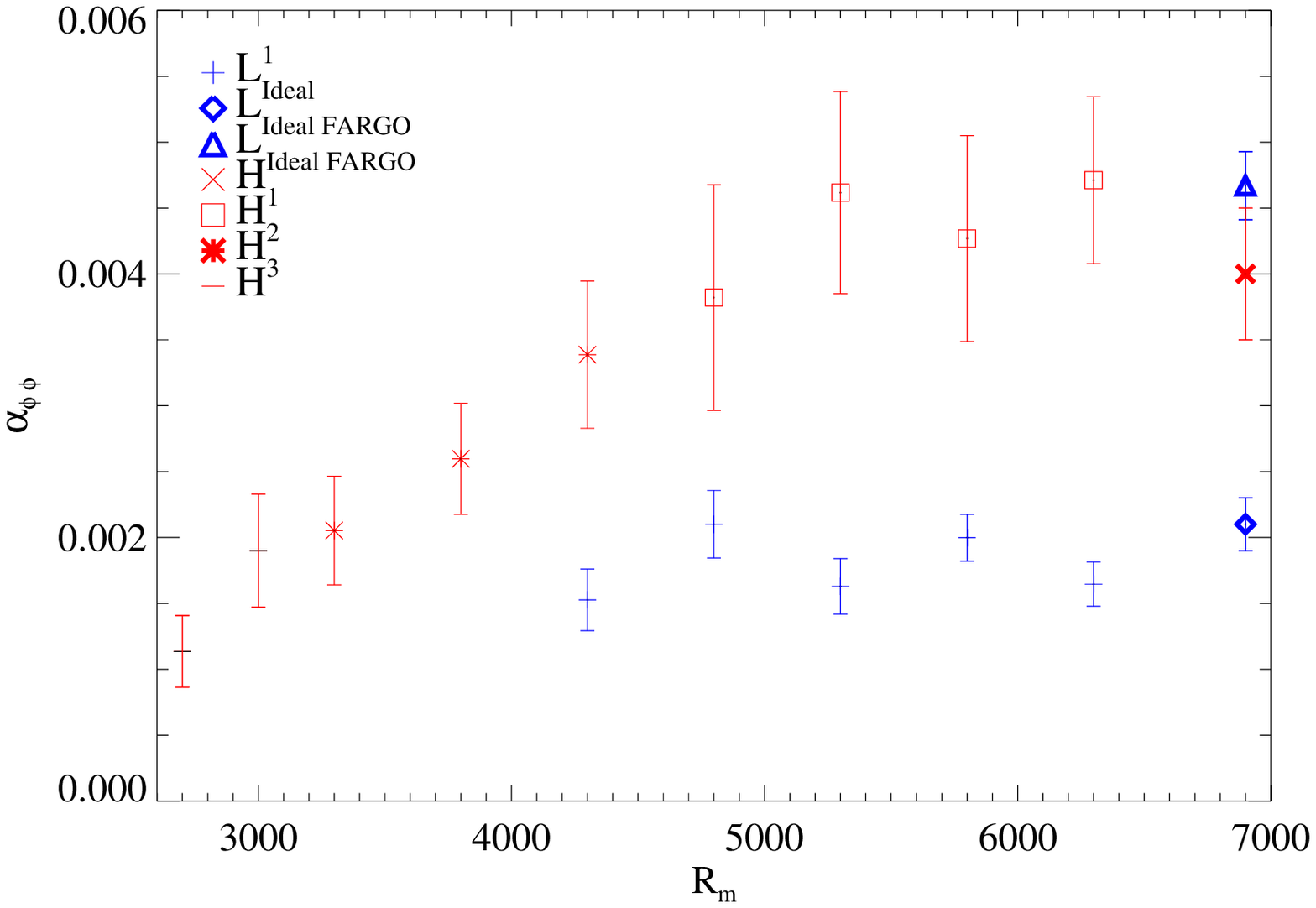,scale=0.57}
\caption{Top: $\alpha_{SS}$ value plotted for different magnetic Reynolds number $R_m$  for a time average of 25 to 60 local orbits.
Middle: Turbulent RMS velocity in units of the local sound speed.
Bottom: Dynamo $\alpha_{\phi\phi}$ 
of the northern hemisphere. 
The L models are plotted in blue the H models are plotted in red. 
The error bars shows the standard deviation in the time average. Values for the corresponding ideal MHD runs are plotted at $Rm = 6900$ as reference. The accretion stress and the turbulent velocities saturate at magnetic Reynolds numbers above 5000. }
\end{figure}

%

\section{Results}
According to Eq. 3, we obtain a specific value of magnetic Reynolds number at each radius, resulting in a specific value of turbulence. To compare the results at different specific radii the timescale of radial mixing is important. Due to the low magnetic dissipation, the timescale of magnetic diffusion is very long compared to the turbulent mixing timescale. 
The maximum radial mixing scale for an alpha value of 0.01 after 1000 years at the outer radius at 8 AU is around $\delta x=\sqrt{2 \alpha_{SS} c_s H t} \sim 0.5 AU$. 
For our analysis we use specific radial positions which are separated enough. In chapter 3.5 we will investigate how the radial transport of magnetic field effects the local evolution.
The $\alpha_{SS}$ value at a specific radial position is calculated with
\begin{equation}
\rm \alpha_{SS}(r) = \left \langle \frac{ \int_r \rho \Bigg(
\frac{v'_{\phi}v'_{R}}{c^2_s} -
\frac{B_{\phi}B_{R}}{4 \pi \rho c^2_s}\Bigg)dV} {\int_r \rho dV} \right \rangle.
\end{equation}
The integral is done for each radius separately $\int_r dV = \int_{r_l}^{r_r} \int_{\theta} \int_{phi} r^2 \sin{\theta} dr d\theta d\phi$ with $r_l$ and $r_r$ is the left and right radial cell boundary. The $\alpha_{ss}$ is mass weighted to recover the correct total value. 
In chapter 3.1 we compare results having the same local time period. Here we average $\alpha_{SS}(r)$, the turbulent velocities 
$V_{RMS}=\sqrt{(\delta v_r)^2 +(\delta v_\theta)^2 +(\delta v_\phi)^2}$ and the dynamo $\alpha_{\phi \phi}$ between 25 and 60 local orbits. 
This ensures that we compare the same dynamical evolution at a specific radius and Reynolds number.
The space average of $V_{RMS}$ and $\alpha_{\phi \phi}$ is done at the midplane region between $0 - 1.5$ scale heights.
In chapter 3.2 we concentrate on the longterm evolution for different values of $R_m$.
There we include the two MRI criteria, the Elsasser number
\begin{equation}
\Lambda = \frac{B_z^2}{\rho \eta \Omega}
\end{equation}
as well as the factor Q
\begin{equation}
Q = \frac{\lambda_c}{\Delta \phi}=2 \pi \sqrt{\frac{16 B_\phi^2}{15 \rho c_s^2} } \frac{H}{R \Delta \phi}. 
\end{equation}
The Q factor is the ratio of the MRI fastest-growing azimuthal wavelength to the azimuthal cell size. The value should be larger than 8 \citep{flo10}. In chapter 3.3 we concentrate on the run $L^1$ having the largest dead-zone. Here we present results of the dominant hydro-dynamical motions.


\subsection{Time averaged statistics}
Due to the change of the rotation period $\Omega$ with radius, the comparison of the turbulence statistics at different radii is limited 
to the number of rotations at the outer radius. We therefore present the comparison for the different Reynolds numbers 
between a time average of $25-60$ local orbits. We note that in this period the initial net magnetic flux has already vanished and we have a zero-net flux MRI turbulence.
We note also that the results from model $L^1$ should be excluded. They indicate that here the explicit dissipation is not resolved and the numerical dissipation dominates.
We still overplot the results obtained by this lower resolution (L models, Fig. 1 blue color) to estimate the total magnetic Reynolds number. This model will be still useful as it shows very 
similar conditions as present in dead-zones.\\
Fig. 1 combines the results of $\alpha_{SS}$, the turbulent RMS velocity $V_{RMS}$ and the dynamo $\alpha_{\phi \phi}$ value as a function of the magnetic Reynolds number.
The results show for all turbulent quantities a saturation for magnetic Reynolds numbers around 5000 and above. 
Below $R_m = 5000$ the MRI is damped and the turbulence decreases.
We observe a slightly higher turbulence level for high magnetic Reynolds numbers compared to fully ideal MHD run.\\
There is a saturation of $\alpha_{SS}$ at around $0.015$ value for magnetic Reynolds numbers greater than 5000, see Fig. 1 top.
Below $R_m < 5000$, $\alpha_{SS} $ drops down to $0.005$ at around $R_m \sim 3300$. Even one could see a flattening at around $R_m = 3300$ the values are still dominated by Maxwell stress which indicates that the MRI is still operating. 
The turbulent velocity scales roughly with the square root of $V_{RMS} \sim \sqrt{\alpha}c_s$, found in recent local box simulations of dead-zones \citep{oku11}.
We observe a saturation around $V_{RMS} = 0.15 c_s$ (Fig. 1, middle). Below $R_m < 5000$, $V_{RMS} $ drops down to $0.09$ at around $R_m \sim 3000$. We note again that these values represent the turbulent motions at the midplane, in the corona the turbulent velocity increases also in the dead-zone models.
%
The dynamo action could be expressed by showing the correlations $\overline{EMF'_\phi} = \alpha_{\phi \phi} \overline{B_\phi}$ with 
$EMF'_\phi = v'_rB'_\theta - v'_\theta B'_r$. We follow the analysis described in chapter 3.7 in \citep{flo11b} and 
$\alpha_{\phi\phi}$ is normalized over $c_s$. The dynamo action in the northern hemisphere is shown in Fig. 1, bottom. $\alpha_{\phi\phi}$ saturates at around 0.004 and decreases only down to 0.002 at $R_m = 3000$.
In all runs the sign of $\alpha_{\phi\phi}$ in the southern hemisphere is negative with similar amplitude as the corresponding value in the northern hemisphere.\\
The results of the ideal MHD runs show convergence. Here the FARGO MHD method plays an important role in decreasing numerical dissipation as it is 
presented in Mignone et al. 2012, A\&A accepted. $\alpha_{SS}$ converge around 0.01.
We note again that the results presented in Fig. 1 reflect mainly the turbulence level at the midplane. In addition we see here no decay of the MRI turbulence due to the short time period.
In the next chapter we will concentrate on the longterm evolution over height for specific Reynolds number.
\subsection{Longterm evolution}
The previous chapter has shown the transition to saturated turbulence in well ionized disk regions.
In this chapter we focus on the longterm evolution, including the turbulence over height. For each model we perform the analysis at a specific Reynolds number which are located in the middle of the domain.
Here, $H^1$ has $R_m = 5500$, $R_m = 4300$ for model $H^{2}$ and $R_m = 3300,3000$ for model $H^3$. In model $L^1$ we expect to have a Reynolds number around $R_m \le  3000$ by comparison with model $H^3$.\\ 
We plot the time evolution of $\alpha_{SS}$ in Fig. 2. 
The analysis show that both ideal runs as well as the resistive runs down to $R_m = 4300$ ($H^1$ and $H^2$), 
show a steady turbulent evolution with $\alpha_{SS} \sim 0.01$.
The models $H^3$ and $L^1$ with $R_m \le 3000$ show a steady decrease of the turbulence until they oscillate 
in the range between $10^{-4} - 10^{-3}$. Due to the decrease we note that the results of the models $H^3$ and $L^1$ plotted in Fig. 1 depend on the averaged time period. For these models, an extended simulation runtime was needed. 
We mark the regions with dominating Reynolds stress in thick (blue and green solid lines). 
The plots show only a short-term domination which indicate that the total integral over height suggest still a operating MHD turbulence. The configuration of the Maxwell stress $\alpha_M=B'_R B'_\phi/4\pi \rho c_s^2$ over height is presented in Fig. 3.
After around 200 inner orbits (years) the initial net azimuthal flux is lost and the flux start to oscillate around zero \citep{fro06,bec11,flo11}. 
For a magnetic Reynolds number above $R_m = 5500$ the ionization is well enough to sustain the MRI turbulence at each height. In the midplane region we found 
a saturation of $\alpha_{SS} \sim 0.01$ at around $R_m \ge 5000$. In model $H^1$, Fig. 3, top, the Elsasser number is between 1 and 10 in the midplane region. 
If the ionization is decreased below $R_m \cong 5000$ the Elsasser number reaches unity at the midplane (see also Fig. 4). 
In model $H^2$, Fig. 3, second from top, there is a small region at the midplane with $0.1 < \Lambda < 1$. The total $\alpha_{SS}$ is still around 0.01 (compare Fig. 2) 
but with higher fluctuations. Locally the magnetic turbulence starts to vanish at the midplane.
At around $R_m \cong 3300$ (model $H^3$) the total $\alpha_{ss}$ gets affected and decreases (compare Fig. 2). 
In model $H^3$ the Elsasser number $\Lambda = 1$ point is around 1 scale height. At the midplane $\Lambda$ drops below 0.1. 
We confirmed the $\Lambda = 0.1$ region equivalent with the $\lambda_{res} = H$ definition by \citet{oku11}, Fig. 1 therein.
At this point also the azimuthal MRI wavelength becomes unresolved ($Q \le 8$, red solid line). At the midplane the Reynolds stress dominate over the Maxwell stress (Fig. 3, yellow solid line).
At $R_m \sim 3000$, the MRI gets strongly damped. For model $L^1$ and model $H^3$ after around 500 years, 
the poloidal magnetic fields decreases until the Elsasser number drops below 0.1 as well as the azimuthal MRI wavelength becomes unresolved (Fig. 3, red solid line). If the turbulence is able to survive depends also on the surrounding layers. 
Even we choose a constant resistivity, the turbulence in the runs $H^3$ and $L^1$ is similar as they are present 
in dead-zones. Turbulent upper layers of the disk are usually active \citep{tur08,dzy10,bai11}. 
In this active layers the turbulent mixing is also stronger and active channels can reach into the dead-zone.
Fig. 4 presents a 2D $r-\theta$ slice of radial magnetic field overplotted with the $\Lambda = 1$ line for model $H^1$ and $H^2$ after 1000 years. 
At around $R_m = 5000$ the dissipation starts to damp the MRI at the midplane while there is still MRI turbulence in the corona of the disk. A similar shape could be expected at 
the outer edge of dead-zones (Dzyurkevich et al. 2012, ApJ subm.).\\
We summarize that for magnetic Reynolds numbers above $R_m > 5000$ we observe a saturated and converged MRI turbulence in zero-net flux stratified simulations down to the midplane region.
The Maxwell stress dominates in each region the Reynolds stress.
Below $R_m < 5000$ the turbulence starts to decay at the midplane. The Elsasser number drops below 1 and there regions with low magnetic fields and dominating Reynolds stress.
At around $R_m \cong 3000$, the Elsasser number drops below 0.1 at the midplane. Long-period oscillations becomes visible with MRI activation and decays. Below $R_m < 3000$ the MRI cannot operate. Here, Reynolds stress dominates over the Maxwell stress. Dependent on the surroundings, the upper layers are still active having MRI turbulence.
The results show that for magnetic Reynolds numbers down to $R_m \cong 4300$, the MRI driven by a zero-net flux azimuthal magnetic field, can sustained the turbulence with $\alpha_{SS} \cong 0.01$. 
Models $H^3$ and $L^1$ with $R_m < 3000$ show a decrease of the turbulence and eventually a dominating Reynolds stress. 
From this results we conclude that the critical magnetic Reynolds number should around $ R^{crit}_m \le 3000$.
\citet{ois11} also found a critical magnetic Reynolds number around $R_m \sim 3000$ in local box simulations.
In the next chapter we concentrate on the run $L^1$ which present similar conditions as they are present in dead-zones.
\begin{figure}
\psfig{figure=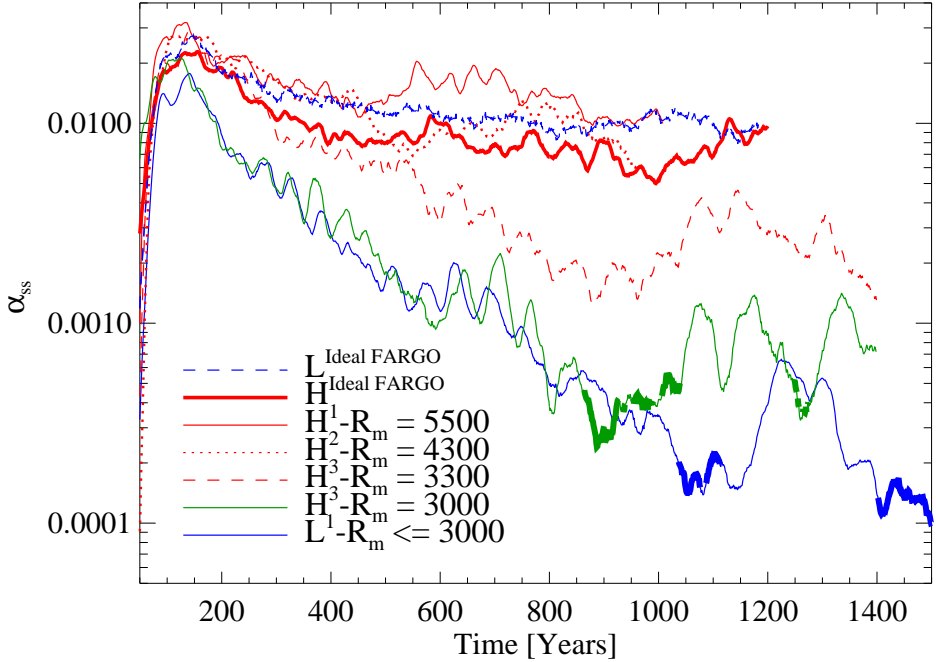,scale=0.80}
\caption{$\alpha_{SS}$ time evolution for different models. The value is taken in the middle of the domains 
for $H^{1}$ at $R_m = 5500$, for model $H^{2}$ at $R_m = 4300$ and $R_m = 3300$ for model $H^{3}$. For model $L^1$ we assume $R_m < 3000$. The ideal MHD models as well as the resistive models down to $R_m = 4300$ show $\alpha$ values around 0.01. We mark the time with dominating Reynolds stress with a thick line.}
\end{figure}

\begin{figure}
\vspace{-0.5cm}
\psfig{figure=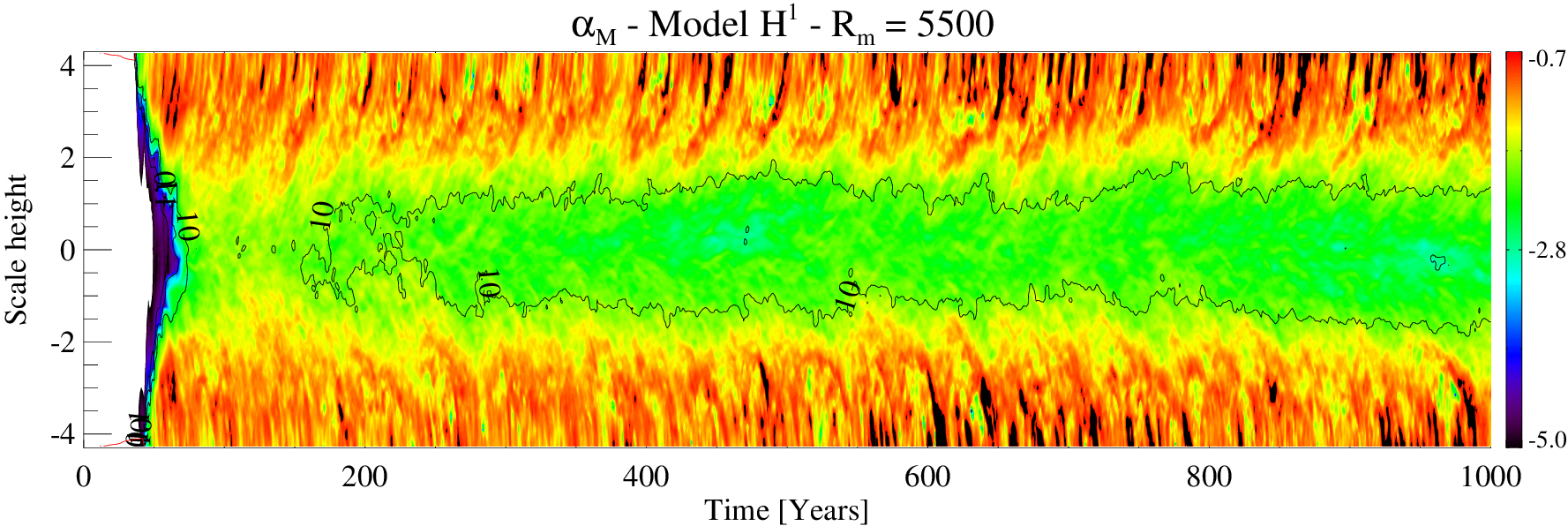,scale=0.55}
\psfig{figure=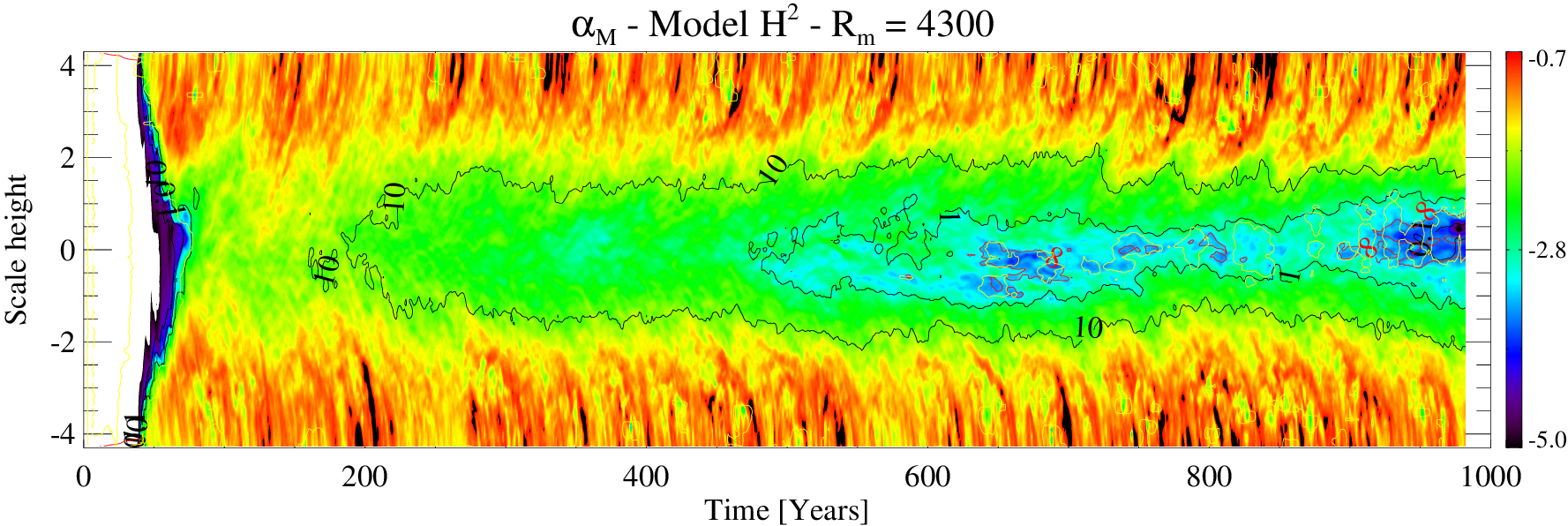,scale=0.55}
\psfig{figure=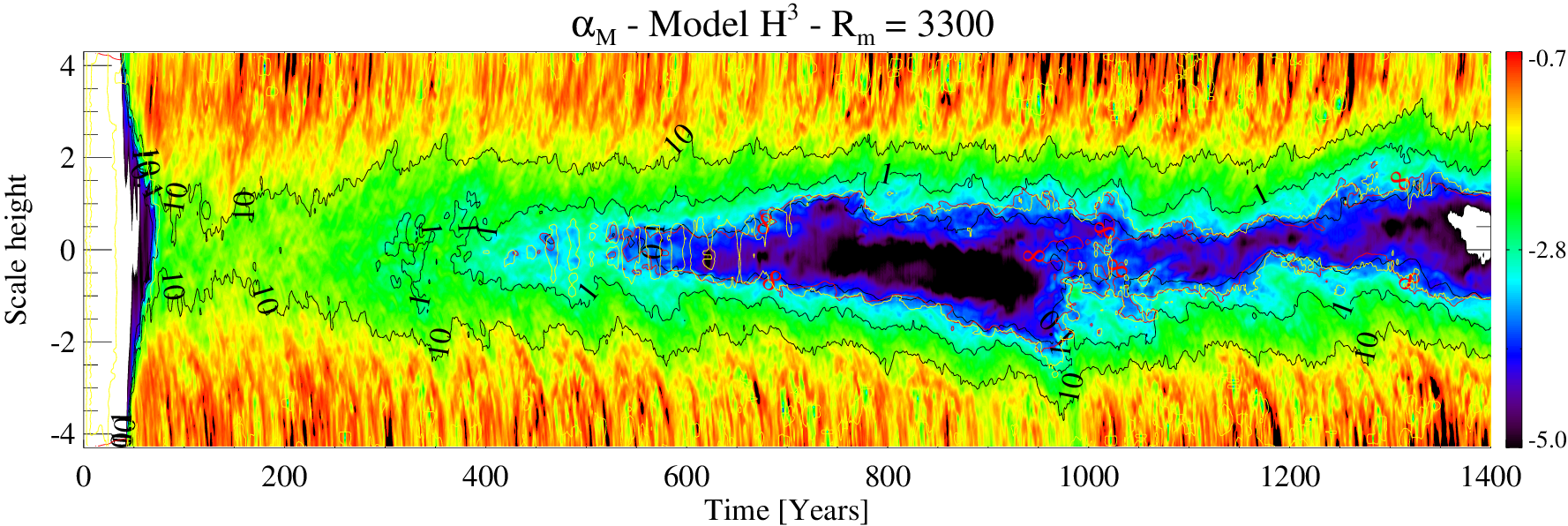,scale=0.55}
\psfig{figure=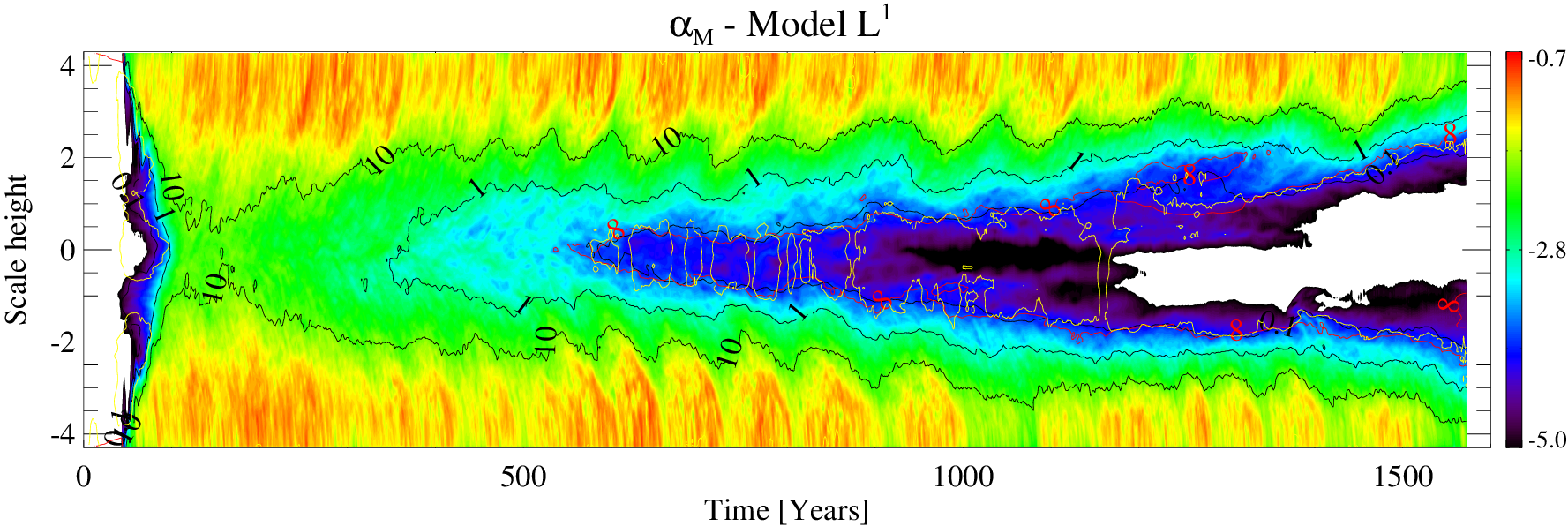,scale=0.55}
\caption{Evolution of the Maxwell stress $\alpha_M$ over height. For model $H^1$ at $R_m = 5500$ (top), $H^2$ at $R_m = 4300$ (second from top), 
$H^3$ at $R_m =3300$ (third from top) and $L^1$ at $R_m \le 3000 $ (bottom).
The black solid lines show the Elsasser numbers 10,1, and 0.1. The red solid line shows $Q = 8$. The yellow line shows 
the region with dominating Reynolds stress.}
\end{figure}

\begin{figure}
\psfig{figure=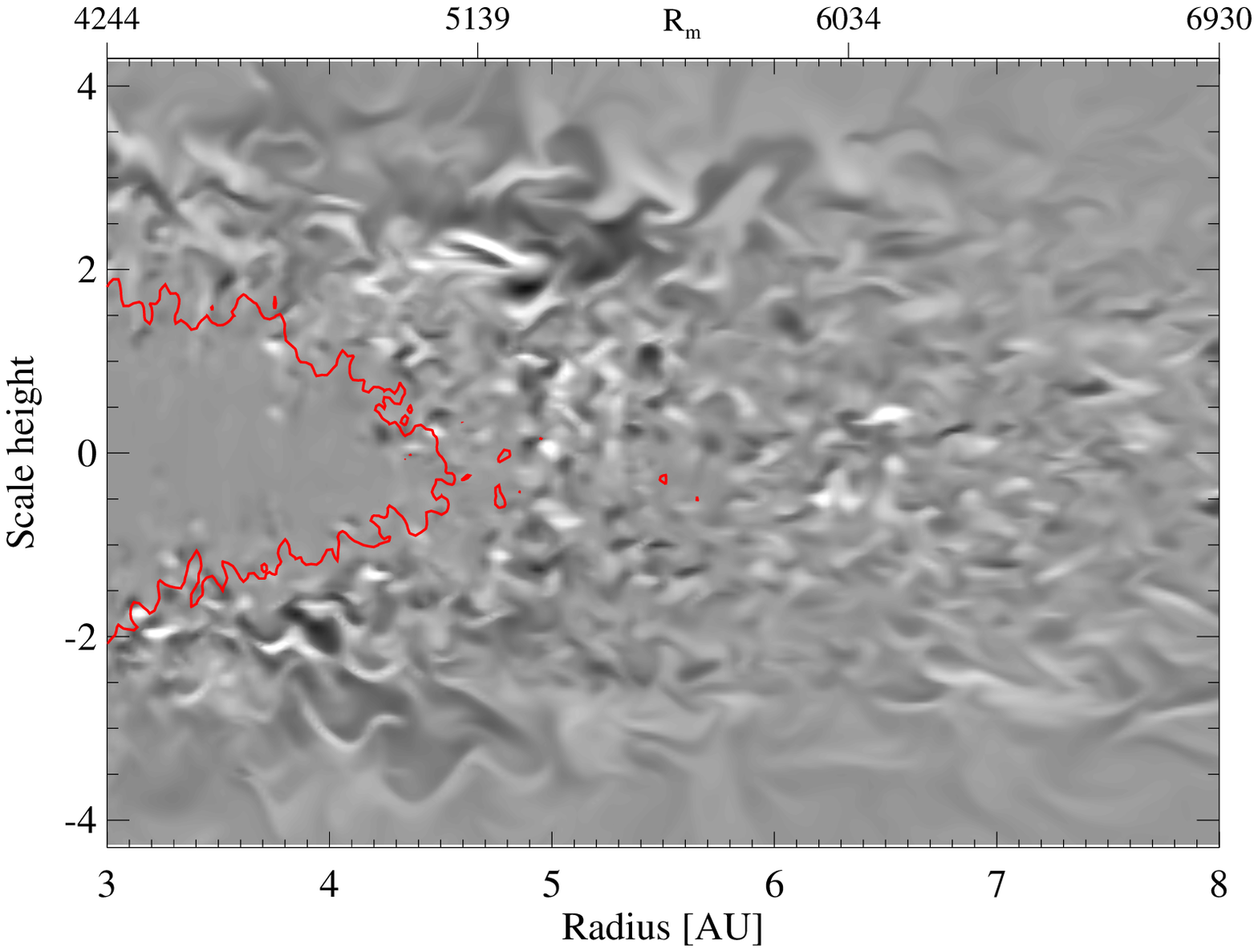,scale=0.80}
\psfig{figure=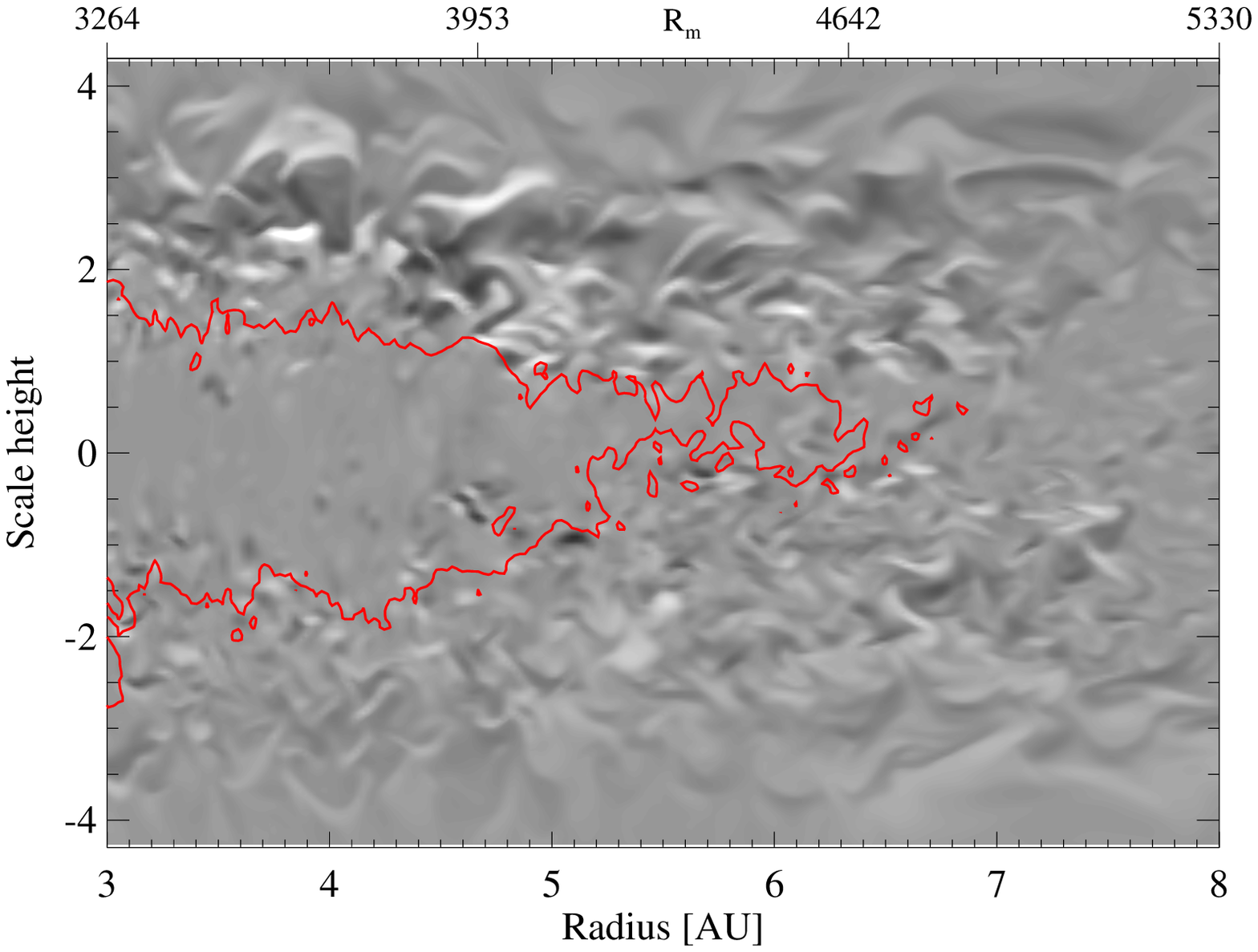,scale=0.80}
\caption{Contour plot of radial magnetic field in the $r-\theta$ plane after 1000 years overplotted with the
Elsasser number $\Lambda = 1$ line. In both models, $H^{1}$ (top) and $H^{2}$ (bottom) the critical 
value of $R_m$ of resolved MRI at the midplane region is around $R_m^{crit} \cong 5000$.}
\end{figure}

\begin{figure}
\vspace{-0.5cm}
\psfig{figure=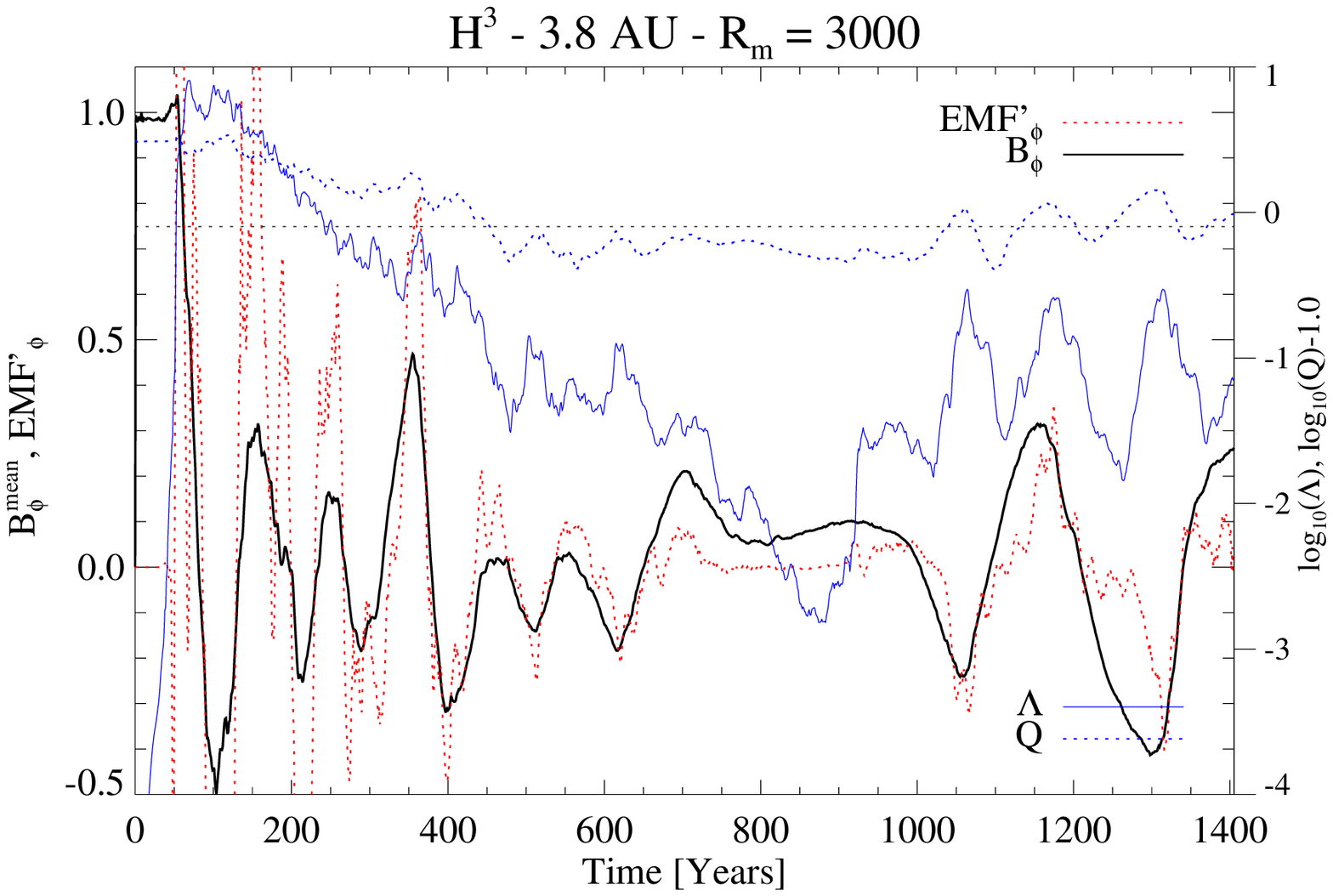,scale=0.80}
\psfig{figure=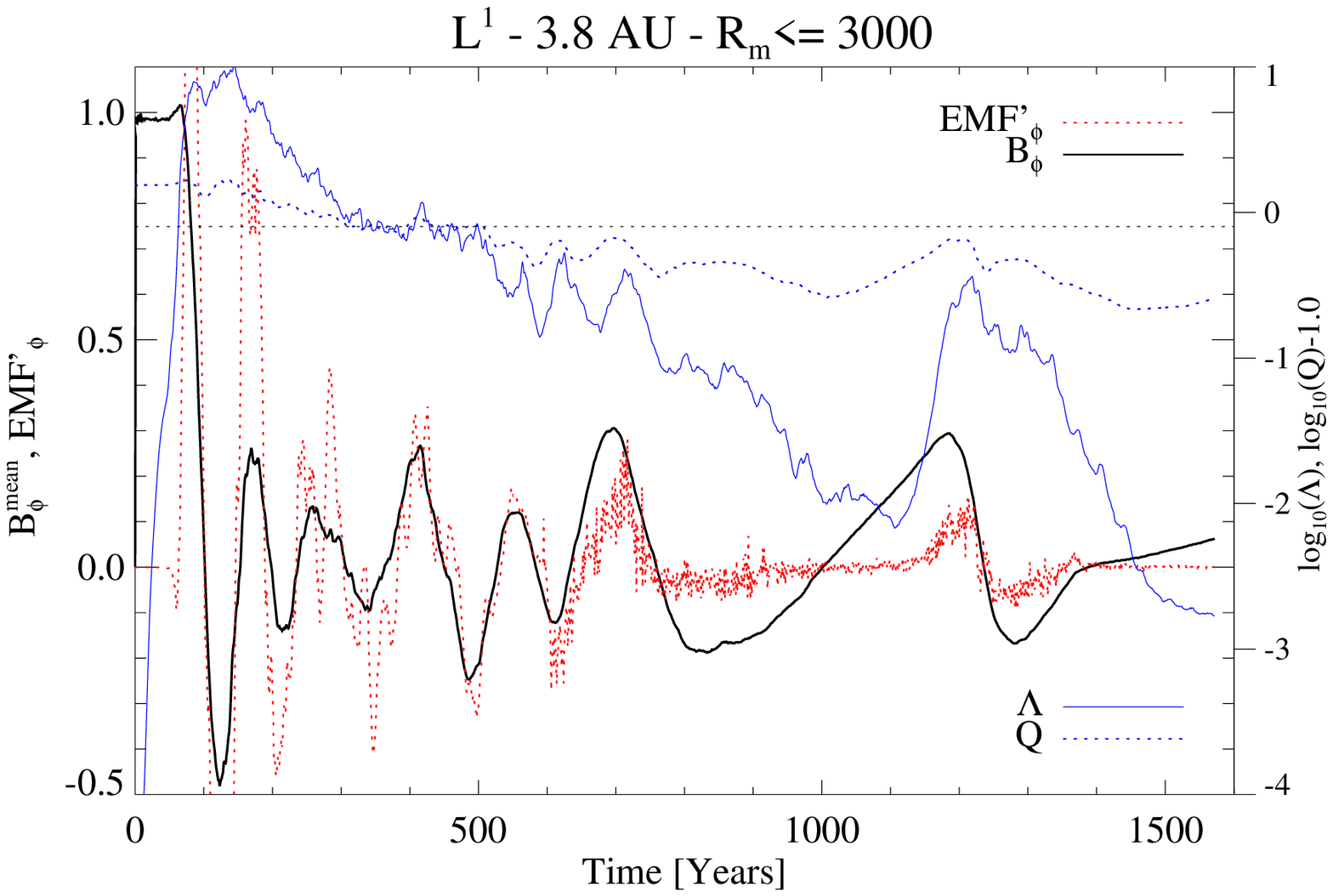,scale=0.80}
\caption{Time evolution of mean azimuthal field (black solid line), turbulent $EMF'_\phi$ (red dotted line), $\Lambda$ (blue solid line) and the Q value (blue dotted line) for model $H^3$ at $R_m = 3000$ (3.8 AU) (top). 
and for model $L^1$ at 3.8 AU (bottom). Values are calculated in the southern hemisphere between (0 - 1.5 scale heights). The black dotted line shows the value of $Q=8$. In model $L^1$ we observe a relaunch of MRI between 1200 and 1300 years in the dead-zone. For model $H^3$ we observe only short-period switch off between 800 and 900 years.}
\end{figure}

\subsection{Dead-zone oscillation}
Before we move to the hydro-dynamical motions in dead-zones we want to investigate the long-term oscillations of the Maxwell stress we observed in the previous chapter. This oscillations could be connected to some sort of dynamo process.
\citet{sim11} presented in local box simulations a sudden transition from a 'low' turbulent state to a 'high' turbulent state. We observe in model $H^3$ and $L^1$ for $R_m \sim 3000$ a similar effect.
The results are combined in Fig 5 showing the mean toroidal magnetic field overplotted with the turbulent $EMF'_\phi$ (times 1000), the Elsasser number and the Q value. 
For this analysis we choose the southern hemisphere between 0 and 1.5 scale heights. 
After around 200 years the initial mean toroidal magnetic field (black solid line) vanishes and starts to oscillate. There is a clear correlation between the turbulent EMF (red dotted line) 
and the mean field which indicates a working $ \alpha \Omega $ dynamo. In both models, after around 
500 years the MRI from azimuthal magnetic fields becomes unresolved ($Q < 8$ blue dotted line below black dotted line). Here the Elsasser number drops below unity which still allows for damped MRI growth.
After 800 years the Elsasser number drops below 0.1. At this stage there is no MRI working and we are in the dead-zone stage. Here also the oscillations of mean field and turbulent EMF stop. In model $H^3$ the MRI switch on again after less then 
200 years (27 local orbits) with a larger oscillation period. In model $L^1$ we observe a late switch on at 1100 years for only 150 years (20 local orbits). Afterwards the Elsasser number reaches nearly $10^{-3}$ and we don't expect a relaunch again. The sudden increase of $\Lambda$ is due to a sudden increase of poloidal magnetic field. 
This sudden increase could be connected with the accumulation of toroidal magnetic fields (black solid line in Fig. 5, bottom) in the dead-zone 
as well as due to radial transport of magnetic field (see chapter 3.5). 
In Fig. 6, we present an azimuthal slice of toroidal magnetic field in the $r-\theta$ plane after 1500 years for model $L^1$. 
In the dead-zone ($\Lambda < 0.1$) we observe locally a Q factor of 8 and above which correspond
to a plasma beta below 50, still the dissipation is too high to relaunch the MRI growth by azimuthal fields. 
%
In general the line of resolved azimuthal field $Q < 8$ matches the one of decaying MRI $\Lambda < 0.1$ (compare Fig. 3).\\
\subsection{Hydrodynamical motions}
This dead-zone region is also interesting in terms of the hydro-dynamical motions.
A close looks at the radial velocity in the $r-\phi$ midplane in Fig. 7 reveals velocity amplitudes around $0.2 c_s$. 
On the first look these hydro-dynamical waves appear to be linear waves with a single mode. But a closer look at the Fourier spectra of the radial velocity reveals a more detailed picture.
We calculated the spectra in the dead-zone region between 4 and 5 AU, time averaged between 1300 and 1500 years in model $L^1$.
In Fig. 8, we compare this spectra with the one obtained in fully ionized disk from model $L^{Ideal\ FARGO}$. 
In the dead-zone there is a peak at $m=6$. The turbulence at lower scales (green solid line) is around 1 order of magnitude below the value in the fully MRI turbulent region.
Beside the density waves we observe anti-cyclonic vortices (Fig. 7 between 6 and 7 AU) which create large extended spiral arms. 
They are produced at around 8 AU where we have a positive density slope due to the buffer zones.
We calculated the relative vorticity $\omega = ((\nabla \times V)_{\theta}  - (\nabla \times  V_{Kepler})_{\theta})/ (\nabla \times v_{Kepler})_{\theta}$ between $-0.8$ and $-0.5$ in the large vortex. 
The growth of vortices at the border of the dead zone can be due to the Rossby wave instability as proposed by \citet{van06} and recently investigated by \citet{lyr12}.
The vortex structure is presented in Fig. 9. The vertical extension is around $\pm 2$ SH, the radial extension is $2H^r \sim 1 AU$ and the azimuthal one is around $10H^\phi \sim 4AU$.
The vortex is dragged by the hydro-dynamical surroundings which gives him the concave (Fig. 9 , top, dominant radial inward velocity), convex (Fig. 9, bottom, $V'_{\phi} > \overline{V}_\phi $ ) shape respectively. The magnetic fields (Fig. 9, bottom black vectors) are not present inside the vortex but in the surrounding layers. We measure a plasma beta of $\sim 10^5$ inside the vortex compared 
to $\beta \sim 100-1000$ in the surrounding region.

\begin{figure}
\psfig{figure=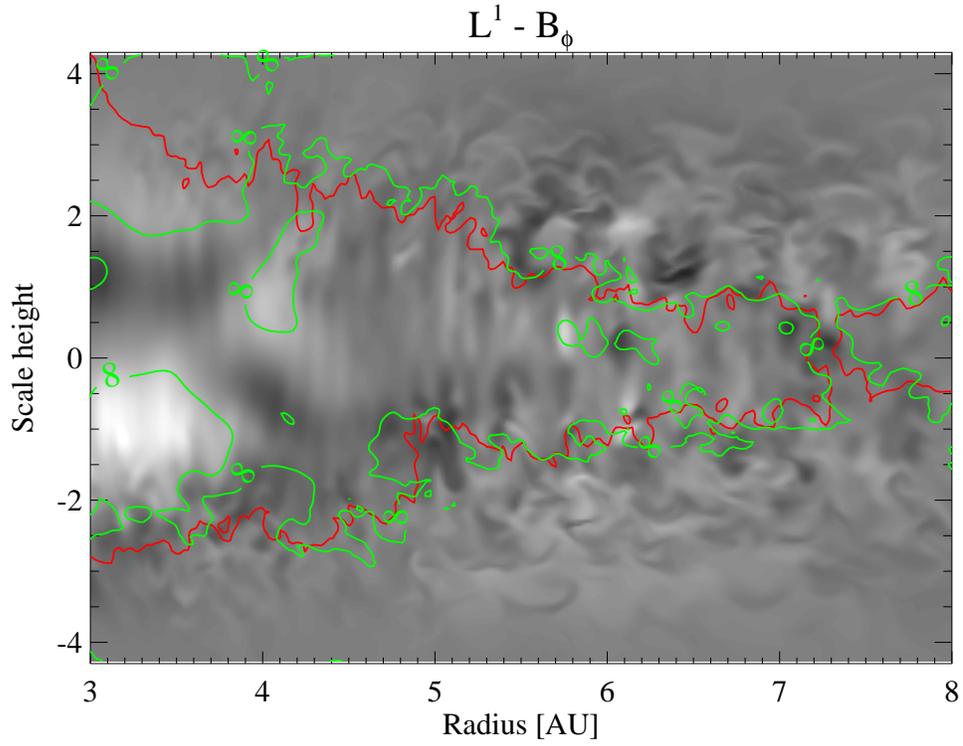,scale=0.80}
\caption{Contour plot of azimuthal magnetic field in the $r-\theta$ plane after 1500 years overplotted with the
Elsasser number $\Lambda = 0.1$ red solid line and the $Q=8$ green solid line. We observe in the dead-zone relative strong (resolved) 
toroidal magnetic fields.}
\end{figure}

\begin{figure}
\psfig{figure=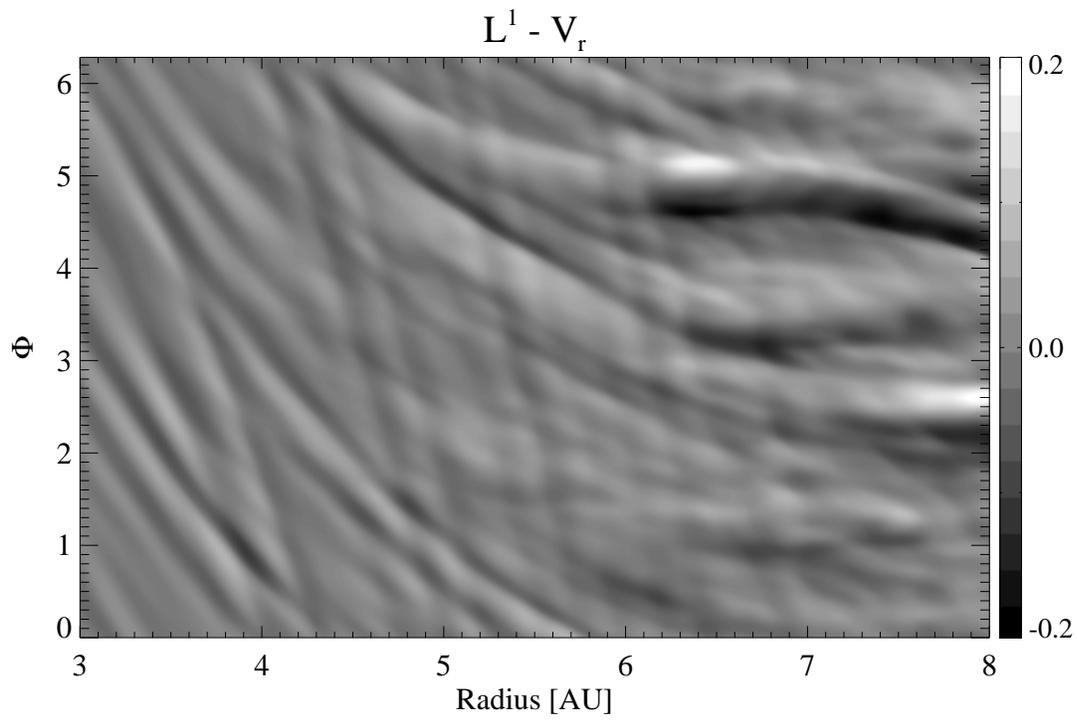,scale=0.80}
\caption{Contour plot of radial velocity in the $r-\phi$ midplane after 1500 years in units of the sound speed. 
The dominating Reynolds stress in the dead-zone is produced by linear waves.}
\end{figure}

\begin{figure}
\psfig{figure=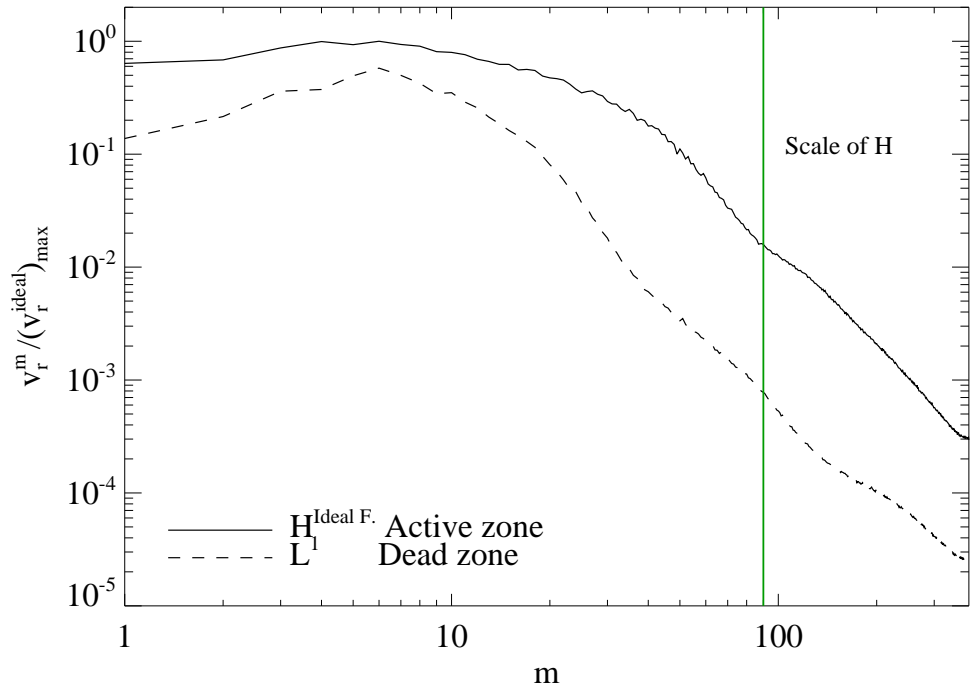,scale=0.80}
\caption{Fourier spectra of radial velocity over azimuthal wavenumber at the midplane compared for the active zone (model $L^{Ideal\ FARGO}$ and the dead-zone (model $L^1$). 
The density wave peak appears in both models at $m=6$. We normalize with the amplitude at $V_r^{Ideal}(6) = 0.25 c_s$.}
\end{figure}

\begin{figure}
\psfig{figure=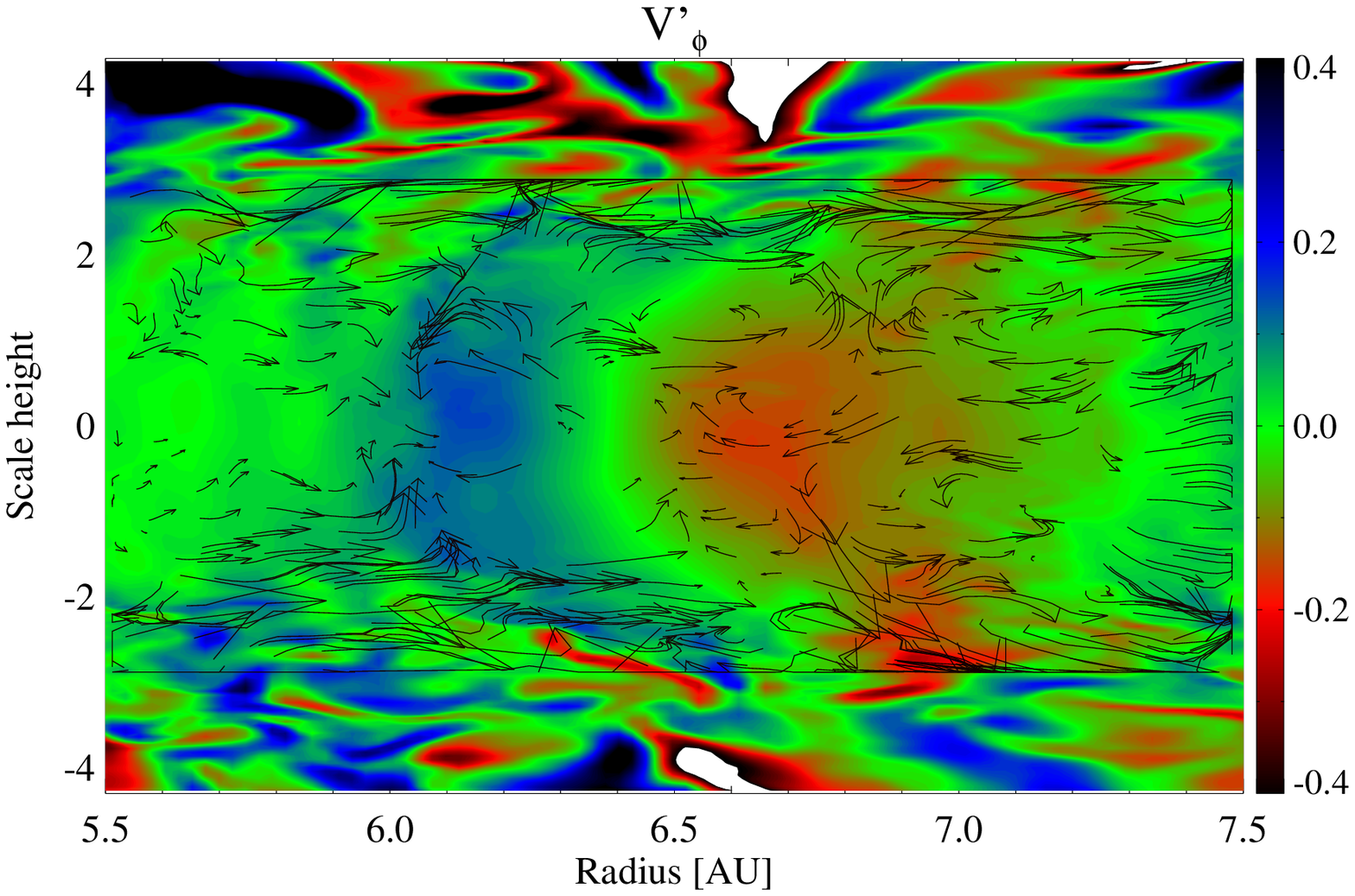,scale=0.80}
\psfig{figure=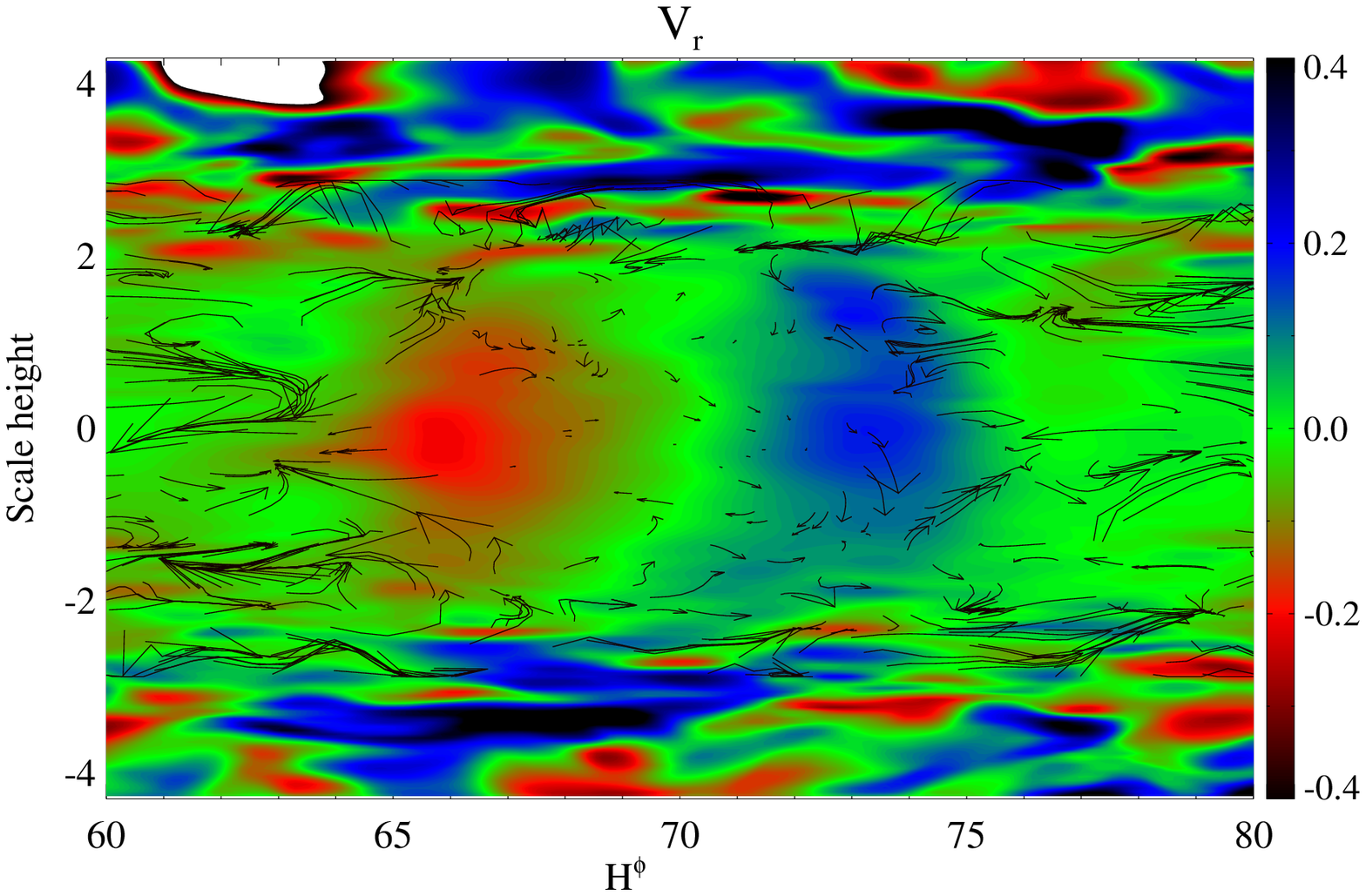,scale=0.80}
\caption{Top: Contour plot inside the vortex of turbulent azimuthal velocity in the $r-\theta$ plane overplotted with the velocity vectors.
Bottom: Contour plot inside the vortex of turbulent radial velocity in the $\phi-\theta$ plane overplotted with the magnetic field vectors.
Color table present the units of the sound speed.}
\end{figure}

\subsection{Radial mixing}
We already mentioned the effect of radial transport due to turbulent mixing. In this subsection we want to focus on the importance of radial transport of magnetic fields. We calculate the divergence of the radial Poynting flux $\int \nabla \cdot S_r dt $ integrated over time for the total $\theta$ and $\phi$ domain at $4.6 AU$ using one scale height in radius. We integrate the divergence of the radial Poynting flux over every output ($dt = 0.1$ local orbits) and compare it with the total magnetic energy. Fig. 10 shows how much magnetic energy is transported at each output (0.1 local orbits). In the fully turbulent layers the net radial transport is around zero with fluctuations of around 5 \%. In the transition zone and the dead-zone the radial transport becomes more important and these fluctuations reach peaks of 10 to 20 \%.

\begin{figure}
\psfig{figure=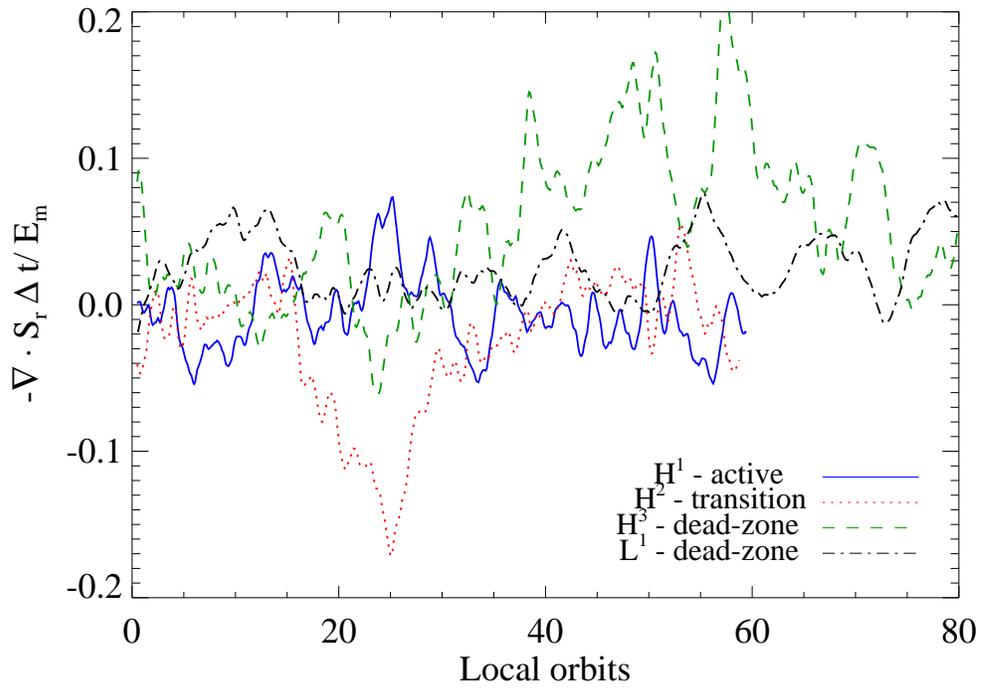,scale=0.80}
\caption{Time integral of divergence of radial Poynting flux over total magnetic energy for model $H^1$ ($R_m = 5500$, blue solid line), $H^2$ ($R_m = 4300$, red dotted line), $H^3$ ($R_m = 3300$, green dashed line) and $L^1$ ($R_m \le 3000$, black dash dotted line). In the transition region the radial mixing becomes more important.}
\end{figure}

\section{Discussion}
\subsection{Turbulence between active and dead-zones}
In our work we showed that the MRI turbulence saturates around $R_m=5000$ with $\alpha_{SS} \sim 0.01$. 
Here the Elsasser number is around $\Lambda \ge 10$. This turbulence level is sustained for Elsasser number values of $1 < \Lambda < 10$. 
Stratified converged ideal MHD local simulations
\citep{dav10} presented similar values of $\alpha_{SS}$. 
Still our results indicate a higher turbulence level using small explicit resistivity. High resolution simulations using explicit resistivity are needed to confirm the convergence of the accretion stress.
In addition, there are mean-field mechanism, which can temporal increase the value of
$\alpha_{SS}$ \citep{flo11b}.\\ 
For magnetic Reynolds numbers around the critical value $R_m \sim 3000 $ we observe long-period oscillations of the accretion stress as well as irregular switch off and on of MRI activity. Such oscillations were also reported by \citet{sim11} in local box simulations. We see indications that these oscillations are trigger by some dynamo process as locally there are accumulations of strong mean toroidal fields in the dead-zone. Also radial transport of magnetic field could here play a role. 
In such configurations the corona is still active while in the midplane the MRI is switched of.
Here the Reynolds stress dominate over the Maxwell stress at $\Lambda < 0.1$.
A similar classification of turbulent regions in proto-planetary disks was done by comparing different heights, 
see also in \citet{oku11}, Fig. 1.
\subsection{Numerical, explicit and turbulent dissipation}
In such turbulent simulations there are 3 kinds of important dissipation sources. The numerical dissipation, due to the finite grid size, the explicit resistivity and the turbulent dissipation.
In model $L^1$ the numerical dissipation plays an important role.  
In addition one has to note that using a uniform grid, the numerical dissipation will be larger at the inner part then in the outer part. Here a logarithmic increasing grid would help.
Then there is of course the explicit dissipation which is included in the induction equation in the code. Here one should be sure that the explicit one is higher than the numerical one.
The turbulent dissipation scales with the level of the turbulence. 
Here it is still unclear how this dissipation process works in detail, see also Fig. 9 in \citet{fro07}. 

\section{Summary}
We performed 3D global ideal and non-ideal MHD simulations to study MRI turbulence in low-ionized proto-planetary disks using 
an initial toroidal magnetic field for a range of magnetic Reynolds numbers. 
With our global simulations we are able to investigate the transition regime which is present between the active and the dead-zone in proto-planetary disks. 
We define 3 different disk regimes dependent on the magnetic Reynolds numbers:
\begin{itemize}
\item Above a Reynolds number of $ R_m \gtrapprox 5000$. 
Here MRI turbulence saturates and is sustained. For this region we find steady and converged $\alpha_{SS}$ values around 0.01 in ideal and non-ideal simulations.
The Maxwell stress dominates down to the midplane region. The turbulent velocities reach maximum values around $V_{RMS} = 0.15 c_s$.
The $\alpha\Omega$ dynamo is operating and we find $\alpha_{\phi \phi} = 0.004$ with a positive sign in the northern hemisphere.
The Elsasser number $\Lambda$ stays above 1 in the midplane.

\item Reynolds number between $ R_m \cong 3000-5000$.
Here, the MRI starts to switch of at the midplane region. The Elsasser number $\Lambda$ drops below 1.
Simulations with $3300 < R_m < 5000$ show a still sustained turbulence, supported by radial transport of magnetic fields. 
For $R_m \cong 3000$ the MRI turbulence starts to switch off. 
We expect the critical magnetic Reynolds number around $R^{crit}_m \lessapprox 3000$ and below.

\item Reynolds number around $ R_m \lessapprox 3000$ and below.
Here, there is no MRI at the midplane anymore. The Elsasser number $\Lambda$ drops below 0.1. The turbulence is dominated 
by the Reynolds stress. We observe long-period oscillations of MRI activity and MRI decay. There is an accumulation of toroidal magnetic fields below $\beta < 50$ in the dead-zone.  The turbulence at the midplane is supported by active channels in the corona which pump sound wave into the midplane region. The velocity spectra at the midplane reveals a drop of one order of magnitude at the scale of H in the dead-zone compared to the fully ionized turbulent region. 
We observe long lived anti-cyclonic vortices in the transition regime, creating spiral arms in the dead-zone. 
Magnetic fields are not present in the inner part of the vortex.
\end{itemize}

We thank Sebastien Fromang for the helpful comments on the global models.
We thank Natalia Dzyurkevich for her suggestions and comments during this work.
We thank also Neal Turner, Satoshi Okuzumi and H\'elo\"ise Meheut for their revision.  
We thank Andrea Mignone for supporting us with the PLUTO code.
The research leading to these results has received funding from the
European Research Council under the European Union's Seventh Framework
Programme (FP7/2007-2013) / ERC Grant agreement n° 258729.
Parallel computations have been performed on the Theo cluster of the Max-Planck
Institute for Astronomy Heidelberg located at the computing center of the Max-Planck Society in Garching.

\bibliographystyle{apj}

\end{document}